\documentclass[11pt,a4paper]{article}
\usepackage{jheppub}
\usepackage{float}
\usepackage{color}

\newcommand{\eq}{\!\! =\!\!}

\makeatletter
\newcommand{\vast}{\bBigg@{4}}
\newcommand{\Vast}{\bBigg@{6}}
\makeatother

\newcommand{\beq}{\begin{equation}}
\newcommand{\eeq}{\end{equation}}
\newcommand{\beqa}{\begin{eqnarray}}
\newcommand{\eeqa}{\end{eqnarray}}

\newcommand{\calL}{\mathcal{L}}
\newcommand{\calK}{\mathcal{K}}
\newcommand{\calO}{\mathcal{O}}
\newcommand{\p}{\partial}

\title{\Huge Spherically symmetric analysis on open FLRW solution in non-linear massive gravity}

\author[\spadesuit\bigstar]{\Large Chien-I Chiang,}
\author[\spadesuit]{\Large Keisuke Izumi}
\author[\spadesuit\heartsuit\diamondsuit\clubsuit]{\Large and Pisin Chen}
\affiliation[\spadesuit]{\large Leung Center for Cosmology and Particle Astrophysics,\\
 National Taiwan University, Taipei 10617, Taiwan}
\affiliation[\heartsuit]{\large Department of Physics, National Taiwan University, Taipei 10617, Taiwan}
\affiliation[\diamondsuit]{\large Graduate Institute of Astrophysics, National Taiwan University, Taipei 10617, Taiwan}
\affiliation[\bigstar]{\large Department of Physics, University of California, Berkeley, CA 94720, U.S.A. }
\affiliation[\clubsuit]{\large Kavli Institute for Particle Astrophysics and Cosmology,\\
SLAC National Accelerator Laboratory, Menlo Park, CA 94025, U.S.A.}

\emailAdd{chienichiang@berkeley.edu}  
\emailAdd{izumi@phys.ntu.edu.tw}
\emailAdd{chen@slac.stanford.edu}

\abstract{We study non-linear massive gravity in the spherically symmetric context. 
Our main motivation is to investigate the effect of helicity-0 mode which remains elusive after analysis of cosmological perturbation around an open Friedmann-Lemaitre-Robertson-Walker (FLRW) universe. 
The non-linear form of the effective energy-momentum tensor stemming from the mass term is derived for the spherically symmetric case. 
Only in the special case 
where the area of the two sphere is not deviated away from the FLRW universe, 
the effective energy momentum tensor becomes completely the same as that of cosmological constant. 
This opens a window for discriminating the non-linear massive gravity from general relativity (GR). 
Indeed, by further solving these spherically symmetric gravitational equations of motion in vacuum to the linear order, we obtain a solution which has an arbitrary time-dependent parameter. 
In GR, this parameter is a constant and corresponds to the mass of a star. 
Our result means that Birkhoff's theorem no longer holds in the non-linear massive gravity and suggests that energy can probably be emitted superluminously (with infinite speed) on the self-accelerating background by the helicity-0 mode, which could be a potential plague of this theory. 
}

\date{\today}

\begin{document}

\maketitle

\section{Introduction}
The theory of massive gravity (see \cite{Hinterbichler:2011tt} for a recent review) has a very
long history dating back to 1939 when Fierz and Pauli \cite{Fierz:1939ix} first wrote down the
action describing a free massive spin-2 particle. Since then
the progress in this area has been made sporadically
and over the past few years this topic has been in a renaissance again because of
the recent consistent construction of the non-linear  massive gravity theory\cite{deRham:2010ik,deRham:2010kj}.

The quadratic form of the massive spin-2 action is uniquely fixed as the linearized general relativity with the Fierz-Pauli mass term \cite{VanNieuwenhuizen:1973fi} by the ghost-free and tachyon-free conditions.
Nevertheless, this linear massive spin-2 theory suffers from the van Dam-Veltman-Zakharov (vDVZ) discontinuity which gives, in the massless limit, the prediction for light bending of 25 percent difference from that of General Relativity (GR) and hence is inconsistent with observational data.
While a massive spin-2 particle has five physical degrees of freedom, namely the helicity-2, helicity-1 and helicity-0 modes, a massless spin-2 particle has only two helicity-2 modes. The vDVZ discontinuity stems from the fact that helicity-0 mode does not decouple with matters even in the massless limit.

Non-linear effects are expected to relieve the inconsistency problem.
Indeed,
the linear analysis is no longer valid inside a certain distance from the source which is called \emph{Vainshtein radius}.
Because the Vainshtein radius goes to infinity as the graviton mass approaches to zero, we can no longer trust the above linear analysis
in the massless limit.
This opens the possibility that non-linearity may \emph{screen} the helicity-0 mode at a scale rendering the theory compatible with observations.

In compensation for screening the helicity-0 mode,
the non-linearity gives rise to another obstacle,
the \emph{Boulware-Deser ghost} \cite{Boulware:1973my}.
Non-linear interactions change the structure of the constraints which in the linear theory eliminates unwanted degrees of freedom generally, hence ghost mode generally appears.
In addition, the mass of the ghost is typically the same order as that of the massive graviton.
We know from the experimental and observational data that the mass of graviton should be small, 
and thus we cannot simply ignore the ghost mode.
The ghost issue has been the core problem of massive gravity for a very long time
and has been actively studied in the past few years 
\cite{Creminelli:2005qk,Izumi:2008st,Izumi:2007pb,Izumi:2007gs,Izumi:2006ca} from the effective field theory approach \cite{ArkaniHamed:2002sp}.

Recently a two parameter non-linear massive gravity which is ghost-free has been developed by de Rham, Gabadadze and Tolley (dRGT) \cite{deRham:2010ik,deRham:2010kj}.
In this construction, coefficients of higher order terms are chosen at each order such that in the decoupling limit the degree of freedom of the BD ghost is removed.
Because of its construction, this theory is potentially free of BD ghost.
Moreover, it was exactly proved that in this theory BD ghost does not show up at fully nonlinear order \cite{Hassan:2011hr,Hassan:2012qv,deRham:2011rn,deRham:2011qq}.
As a result, the form of the nonlinear terms are constrained and the mass and interaction terms are described with only three parameters.

Once a gravitational theory is constructed, it is always interesting to survey its cosmological implication \cite{Gumrukcuoglu:2011ew,Gumrukcuoglu:2011zh,D'Amico:2011jj,Koyama:2011xz,Koyama:2011yg,Comelli:2011wq,Nieuwenhuizen:2011sq,Koyama:2011wx,Sbisa:2012zk,Berezhiani:2011mt,Sjors:2011iv,Brihaye:2011aa,Kobayashi:2012fz,DeFelice:2012mx,Gumrukcuoglu:2012aa,Gratia:2012wt,Volkov:2012cf}.
One of the phenomenologically interesting aspects is that small graviton mass naturally explains the smallness of the cosmological constant which we have observed.
The basic idea is that due to the mass, a Yukawa suppression $e^{-mr}$ comes in and the gravitational force is weakened at large scale.
So even if we have a large underlying cosmological constant, say the one predicted by usual quantum field theory, it is effectively small in terms of gravity.
The accelerating Friedmann-Lemaitre-Robertson-Walker (FLRW) solution of dRGT massive gravity was found in \cite{Gumrukcuoglu:2011ew} where the spatial geometry can neither be flat nor closed \cite{D'Amico:2011jj} but open.

The perturbation theory with general matter around self-accelerating universe solutions was studied in \cite{Gumrukcuoglu:2011zh}. It was found that the kinetic terms of the helicity-0 and helicity-1 modes vanish, while the helicity-2 modes have time-varying mass. This is rather unexpected since usually massive gravity has five propagating degrees of freedom. This also makes it difficult to compare this non-linear massive gravity with GR because the structure of the propagations in the linearized theory of the former is basically the same as that in the linearized theory of the latter except for the helicity-2
part.

The purposes of this paper are to study the difference between non-linear massive gravity and GR in open FLRW background, and to investigate the effect of helicity-0 mode.
As a first step, we consider a spherically symmetric configuration in the vacuum, say outside of a star, but which asymptotically approaches to self-accelerating open FLRW background solution found in \cite{Gumrukcuoglu:2011ew}.
The corresponding solution in the framework of GR is the de Sitter-Schwarzschild solution, which has only one constant parameter, namely the mass of the star.
We obtain the result that for this spherically symmetric setup, unlike in GR,
an arbitrary time-dependent function appears in the solution in linear analysis.

The rest of this paper is organized as follows. In Sec.~\ref{review} 
we  review the non-linear massive gravity model and its self-accelerating solution. 
In Sec.~\ref{sphr} we analyze the spherical symmetric solutions whose asymptotic structure are the same as the self-accelerating solution. 
Sec.~\ref{Sum} is devoted to the summary of this paper and discussion. 
In Appendix~\ref{calmass}, we show the calculation related to the mass terms.
In Appendix~\ref{App2}, the derivation of the perturbative contribution from the Einstein-Hilbert term is shown.

\section{Review of dRGT massive gravity and its self-accelerating solution}
\label{review}
In order to be self-contained and to set up the notations, we briefly review some basic notions of massive gravity theory, the structure of the non-linear massive gravity proposed by de Rham et al. \cite{deRham:2010ik,deRham:2010kj} and the self-accelerating solution found in \cite{Gumrukcuoglu:2011ew}.

One important concept in massive gravity is that a priori background is needed in order to construct the theory.
This seems to be a drawback
as compared to GR which is background-independent and
can be described as geometry.
However, one should not forget that by starting from the theory of free massless spin-2 particle propagating in Minkowski space and consistently adding non-trivial interactions one can arrive Einstein's GR.
(For introduction to GR from this point of view, see chapter three of \cite{Ortin:2004ms}.)
The background-independence of GR is a result of the consistent massless spin-2 theory.
In this sense, the introduction of a priori background in the massive spin-2 theory is not  intentional.

From a practical point of view, since adding a mass term 
by only using the physical metric $g_{\mu\nu}$ is impossible, it is inevitable that
another metric is required for massive gravity \cite{Hinterbichler:2011tt}. 
Particularly, the trace of $g_{\mu \nu}$ with itself only gives a constant.  Hence, in order to introduce square or higher order terms, one requires
another metric to do contractions and to take traces.
The simplest way is to introduce a flat absolute background $\eta_{\mu \nu}$.
Introducing spin-2 particle $h_{\mu\nu}$ on the flat absolute background, the physical metric $g_{\mu \nu}$ is defined as
\begin{eqnarray}
g_{\mu \nu} = \eta_{\mu \nu} + h_{\mu \nu},
\end{eqnarray}
and we assume that all matters feel the physical metric $g_{\mu \nu}$.

Note that $h_{\mu \nu}$ transforms non-covariantly under diffeomorphism and that the diffeomorphism symmetry is broken by introducing the non-dynamical absolute background. Nonetheless, we can construct an equivalent but \emph{covariant} theory by introducing four scalar fields $\phi^a$ (a=0,1,2,3) called \emph{St\"{u}ckelberg fields}. The relation between the St\"{u}ckelberg fields and the physical metric is defined as
\begin{eqnarray}
g_{\mu \nu} = Z_{\mu \nu} + H_{\mu \nu},
\end{eqnarray}
where
\begin{eqnarray}
Z_{\mu \nu} = \eta_{ab}\p_\mu \phi^a \p_\nu \phi^b
\end{eqnarray}
is called the \emph{fiducial metric}.
If we choose the gauge condition where $\phi^a=x^a$, the fiducial metric $Z_{\mu\nu}$ becomes $\eta_{\mu\nu}$ and the field $H_{\mu \nu}$ is reduced to $h_{\mu\nu}$.
Note that since the St\"{u}ckelberg fields transform as scalars under diffeomorphism transformation, the fiducial metric $Z_{\mu \nu}$ and the field $H_{\mu \nu}$ transform covariantly.
One replaces all occurrences of $h_{\mu \nu}$ in
the action with $H_{\mu \nu} = g_{\mu \nu} - Z_{\mu \nu}$, 
and then arrives in the final expression
where all quantities transform covariantly and the diffeomorphism symmetry is restored.

The action of the non-linear massive gravity proposed in \cite{deRham:2010kj} consists of two parts, the usual Einstein-Hilbert term $S_{EH}$ and a graviton mass term $S_{mass}$:
\begin{eqnarray}
S_g&=& S_{EH}+S_{mass},  \\
S_{EH}&=& \frac{M_{pl}^2}{2}\int dx^4 \sqrt{-g} R ,\\
S_{mass}&=& M^2_{Pl} m^2_g \int d^4x \sqrt{-g} (\calL_2+ \alpha_3 \calL_3 + \alpha_4 \calL_4),
\end{eqnarray}
where $R$ is the Ricci scalar constructed with physical metric $g_{\mu\nu}$.
In preparation for constructing the mass term $S_{mass}$, 
we define a tensor ${\cal K}^{\mu}_{\ \nu}$ as 
\begin{eqnarray}
&&{\cal K}^{\mu}_{\ \nu}=\delta^\mu_{\ \nu}-W^\mu_{\ \nu}, \\
&&W^\mu_{\ \alpha}W^\alpha_{\ \nu}= g^{\mu\alpha}\eta_{ab}\partial_\alpha\phi^a
\partial_\nu\phi^b \equiv Z^{\mu}_{\ \nu} = g^{\mu \rho} Z_{\rho \nu},
\end{eqnarray}
where $\mu, \nu = 0, \cdots,3$ and $a,b =0 , \cdots,3$.
In order to
avoid the appearance of BD ghost, the mass term $S_{mass}$ is constructed as \cite{deRham:2010kj}
\begin{eqnarray}
&&\calL_2=\frac{1}{2}\left(\left[ {\cal K} \right]^2-\left[ {\cal K}^2 \right]\right), 
\label{L_2}  \\
&&\calL_3=\frac{1}{6}\left( \left[ {\cal K} \right]^3-3\left[ {\cal K} \right]
\left[ {\cal K}^2 \right] +2 \left[ {\cal K}^3 \right]\right), 
\label{L_3}  \\
&&\calL_4= \frac{1}{24}\left(\left[ {\cal K} \right]^4 -6\left[ {\cal K}^2 \right]\left[ {\cal K} \right]^2
+3\left[ {\cal K}^2 \right]^2+8\left[ {\cal K} \right]\left[ {\cal K}^3 \right]
-6\left[ {\cal K}^4 \right] \right),
\label{L_4}
\end{eqnarray}
where the square brackets denote trace operation, for example $[\calK] = {\calK^\mu}_\mu$.
Indices are raised and lowered with respect to the physical metric $g_{\mu \nu}$, unless specified otherwise. In terms of ${W^\mu}_\nu$ and ${Z^\mu}_\nu$, $\calL_2$, $\calL_3$ and $\calL_4$ are expressed as
\begin{eqnarray}
\calL_2&=&6-3[W]+\frac{1}{2}[W]^2 -\frac{1}{2}[Z],
\label{L2} \\
\calL_3&=&4-3[W]+[W]^2-\frac{1}{6}[W]^3-[Z]+\frac{1}{2}[W][Z]-\frac{1}{3}[WZ],
\label{L3} \\
\calL_4&=&1-[W]+\frac{1}{2}[W]^2-\frac{1}{6}[W]^3+\frac{1}{24}[W]^4 \nonumber \\
       &&-\frac{1}{2}[Z] +\frac{1}{2}[W][Z] -\frac{1}{4}[W]^2[Z]+\frac{1}{8}[Z]^2-\frac{1}{3}[WZ]+\frac{1}{3}[W] [WZ]-\frac{1}{4}[Z^2].
\label{L4}
\end{eqnarray}
The gravitational equation can be obtained by the variation of the action with respect to the metric; 
\begin{eqnarray}
\frac{M_{Pl}^2}{2}G_{\mu\nu}+ \frac{\delta}{\delta g^{\mu\nu}}S_{mass}=0,
\label{gravitationaleq}
\end{eqnarray}
where $G_{\mu\nu}$ is the Einstein tensor, id est, $G_{\mu\nu}=R_{\mu\nu}-(R/2) g_{\mu\nu}.$

According to paper \cite{Gumrukcuoglu:2011ew}, we show the self-accelerating open FLRW solution in this theory.
With the aim of finding FLRW solution, we fix the gauge such that the St\"{u}ckelberg fields are in the form of open chart of Minkowski spacetime :
\begin{eqnarray}
\phi^0_{(0)} &=& f(t) \sqrt{1+|K|(x^2+y^2+z^2)},\label{phi0} \\
\phi^1_{(0)} &=& \sqrt{|K|}f(t)x,  \\
\phi^2_{(0)} &=& \sqrt{|K|}f(t)y,  \\
\phi^3_{(0)} &=& \sqrt{|K|}f(t)z.
\end{eqnarray}
In this gauge the fiducial metric respects the symmetry of the open FLRW space time:
\begin{eqnarray}
Z_{\mu \nu } = - (\dot f)^2\delta^0_\mu \delta^0_\nu+ |K|f^2\Omega_{ij}\delta^i_\mu \delta^j_\nu,
\end{eqnarray}
where
\begin{eqnarray}
\Omega_{ij}dx^idx^j= dx^2 + dy^2 + dz^2 -
                                  \frac{|K|(xdx +ydy + zdz)^2}{1+|K|(x^2+y^2+z^2)}.
\end{eqnarray}
As for the physical metric, one considers an open ($K<0$) FLRW ansatz;
\begin{eqnarray}
g_{\mu\nu}dx^\mu dx^\nu = -dt^2 + a^2(t) \Omega_{ij}dx^idx^j.
\label{FLRW}
\end{eqnarray}
Substituting (\ref{phi0})-(\ref{FLRW}) into the action and varying with respect to $f(t)$ one obtains the solution for the St\"{u}ckelberg fields, which has two branches of solution,
\begin{eqnarray}
f= \frac{a}{\sqrt{|K|}}X_{\pm}, \quad \quad X_{\pm} \equiv \frac{1+2\alpha_3 + \alpha_4 +\pm \sqrt{1+\alpha_3 + \alpha_3^2-\alpha_4}}{\alpha_3+\alpha_4}.
\label{f}
\end{eqnarray}
When $\alpha_3$ and $\alpha_4$ are the orders of a small parameter $\epsilon$, we have
\begin{eqnarray}
X_+ &=& \frac{2}{\alpha_3 + \alpha_4} + \frac{5\alpha_3 + \alpha_4}{2 (\alpha_3 + \alpha_4)} + \calO(\epsilon) + \cdots ,\\
X_- &=& \frac{3}{2} + \calO(\epsilon) + \cdots.
\end{eqnarray}
The Friedmann equation obtained from this theory is basically that obtained by GR but with cosmological constant replaced by the graviton mass. In particular,
\begin{eqnarray}
H^2 &=& \frac{\Lambda_{\text{eff}}}{3} + \frac{|K|}{a^2},\\
\dot{H}&=&- \frac{|K|}{a^2},
\end{eqnarray}
where 
\begin{eqnarray}
\Lambda_{\text{eff}}= c_{\pm} m_g^2, \label{Lambdaeff}
\end{eqnarray}
\begin{eqnarray}
c_{\pm} = - \frac{\left(1+\alpha_3\pm\sqrt{1+\alpha_3 +\alpha_3^2-\alpha_4} \right)}{(\alpha_3 + \alpha_4)^2}\left( 1+\alpha_3^2-2\alpha_4\pm(1+\alpha_3)\sqrt{1+\alpha_3 +\alpha_3^2-\alpha_4}\right).
\end{eqnarray}
In the small parameter limit ($\alpha_3,\alpha_4 = O(\epsilon)$ ), $c_{\pm}$ becomes
\begin{eqnarray}
c_+ &=& - \frac{4}{(\alpha_3 + \alpha_4)^2}-\frac{6(\alpha_3-\alpha_4)}{(\alpha_3 + \alpha_4)^2}- \frac{3(3\alpha_3-\alpha_4)^2}{4(\alpha_3+\alpha_4)^2} + \calO(\epsilon)+\cdots ,\\
c_- &=& \frac{3}{4} + \calO(\epsilon).
\end{eqnarray}
Note that the ``positive" branch of solution becomes singular when $\alpha_3$ and $\alpha_4$ go to zero, while the ``negative" branch remains regular.

The Friedmann equations of non-linear massive gravity are the same as GR with cosmological constant, where the mass of the graviton plays the role of effective cosmological constant.
This opens a window for explaining the smallness of the cosmological constant.
Although 
we still need the small parameter,
but the physical meaning is different.
If we regard vacuum energy as the origin of cosmological constant, naive approximation using quantum field theory is of $10^{120}$ order difference from observational results.
Apparently, our knowledge about vacuum is very far from complete. On the other hand, because of the Yukawa suppression, rendering graviton small mass naturally explains the accelerated expansion no matter how large the vacuum energy density is. In fact, one cannot distinguish the contribution of the graviton mass and cosmological constant at the background evolution level. Specifically, if we add a cosmological constant term in the non-massive gravity action, one will arrive at Friedmann equations with a cosmological constant in addition to the graviton mass term which plays the same role as the former. Of course, this does not solve the whole problem but gives another physical interpretation. The quest for the true nature of vacuum still remains to be a very important problem in physics.

\section{Spherically symmetric analysis around self-accelerating Background}
\label{sphr}
It was found in \cite{Gumrukcuoglu:2011zh} that for open FLRW background the helicity-0 and helicity-1 modes have vanishing kinetic terms.
Hence, in the linear analysis, the structure of the propagations of the cosmological perturbation theory on the accelerating background of dRGT massive gravity is the same as that of GR except that the helicity-2 mode in the former has a time-varying mass. Some natural questions then arise: 
Are there other observational consequences where we can see the contribution of helicity-0 mode? How about the helicity-1 mode? What will we find if we go beyond linear analysis to non-linear regime? In this work we will try to address the first and the third
questions.

Since in dRGT massive gravity five degrees of freedom can propagate on Minkowski background, the situation that the helicity-0 and helicity-1 modes do not propagate on an open FLRW background at linear regime is very likely to change if we consider a background that is slightly different from open FLRW.
One example is considering a star in an open FLRW universe.
At a certain distance from the star, where linear perturbation is available but not far enough to neglect its gravitational effect, the helicity-0 mode propagates hence gives an additional force. The spacetime would no longer be de Sitter-Schwarzschild solution, as in the same configuration but with GR as the gravitational theory.
Investigating this kind of spherically symmetric case which asymptotically approaches to open FLRW solution can not only give a strong phenomenological constraint on dRGT massive gravity, but also give us a better understanding on the helicity-0 mode.

A spherically symmetric metric can be generally written as
\begin{eqnarray}
ds^2=-e^{2\Phi}dt^2 + \frac{a^2}{1-Kr^2} e^{2\Psi} \left(dr + \beta dt\right)^2 
+a^2 r^2 e^{2E} d\Omega_{(2)}^2,
\label{metric}
\end{eqnarray}
where $\Phi$, $\Psi$, $\beta$ and $E$ are the functions of $t$ and $r$, 
\begin{eqnarray}
d\Omega_{(2)}^2 = d \theta^2 + \sin^2 \theta d\psi^2,
\end{eqnarray}
and 
for $\Phi=\Psi=\beta=E=0$
this metric becomes exactly the open FLRW metric~(\ref{FLRW}) which is obtained in \cite{Gumrukcuoglu:2011ew}.
Throughout the calculation we will fix the gauge such that the fiducial metric remain diagonal form to all order,
\begin{eqnarray}
Z_{\mu\nu} dx^\mu dx^\nu =-(\dot{f}(t))^2 dt^2+ \frac{|K|(f(t))^2}{1-Kr^2}dr^2
+ |K|(f(t))^2 r^2 d\Omega_{(2)}^2,
\end{eqnarray}
where $f$ is defined in eq.(\ref{f}). 
Gravitational equations of motion can be obtained by variations with respect to $g^{\mu\nu}$. 
Although variations with respect to St\"{u}ckelberg fields give equations of motion, 
all of them can be reduced to the equations of motion from variations with respect to $g^{\mu\nu}$~\cite{Hassan:2011vm}.
We show the contributions of the gravitational equations of motion from 
the mass term and the Einstein-Hilbert term separately. 
After that, we combine them and solve the linearized equations. 

\subsection{Contributions from the mass term} 
\label{masscontribution}
 
In this subsection, we show the contributions from the mass term.
The variations of $[W]$, $[WZ]$ and $[Z^2]$ with respect to $g^{\mu\nu}$ can be written as 
\begin{eqnarray}
&&\!\!\!\!\!\!
\frac{\delta [W]}{\delta g^{\mu\nu}}\equiv{\cal W}_{\mu\nu}
= \frac{\sqrt{\det (\mathbb{Z})}}{2 {\mathbb{W}}_T} \bar g_{\mu\nu} 
+\frac{1}{2{\mathbb{W}}_T}\bar Z_{\mu\nu} + \frac{1}{2}X_\pm e^{-E} (g_{\mu\nu}-\bar g_{\mu\nu}),\\
&&\!\!\!\!\!\!
\frac{\delta [WZ]}{\delta g^{\mu\nu}} \equiv\left({\cal W} Z\right)_{\mu\nu}
=-\frac{3}{2}\frac{\det (\mathbb{Z})}{{\mathbb{W}}_T} \bar g_{\mu\nu}
+\frac{3}{2{\mathbb{W}}_T}\left(\mathbb{W}_T^2-\sqrt{\det Z}\right)\bar Z_{\mu\nu} 
+ \frac{3}{2}X_\pm^3e^{-3E} (g_{\mu\nu}-\bar g_{\mu\nu}), \\
&&\!\!\!\!\!\!
\frac{\delta [Z^2]}{\delta g^{\mu\nu}} =2 Z_{\mu\alpha}Z^\alpha_{\ \nu}
=-2\det (\mathbb{Z}) \bar g_{\mu\nu}
+2\left(\mathbb{W}_T^2-2\sqrt{\det Z}\right)\bar Z_{\mu\nu} 
+ 2X_\pm^4e^{-4E} (g_{\mu\nu}-\bar g_{\mu\nu}),
\label{Z2}
\end{eqnarray}
where 
\begin{eqnarray}
&&\det(\mathbb{Z})= (g^{00}g^{rr}-g^{0r}g^{0r})Z_{00}Z_{rr}, \\
&&\mathbb{W}_T = \sqrt{g^{00}Z_{00}+g^{rr}Z_{rr}+2\sqrt{\det(\mathbb{Z})}} \ ,
\end{eqnarray}
are the determinant of the upper-left  $2\times2$ part of ${Z^\mu}_\nu$ and 
the trace of the upper-left $2\times2$ matrix of ${W^\mu}_\nu$ respectively, the upper-left  $2\times2$ parts of $\bar g_{\mu\nu}$ and  $\bar Z_{\mu\nu}$ are the same as those of ${g^\mu}_\nu$ and ${Z^\mu}_\nu$ respectively, and the other components of  $\bar g_{\mu\nu}$ and  $\bar Z_{\mu\nu}$ are zero\footnote{
Here, our coordinates are $x^\mu=(t,r,\theta, \psi)$
}.
Namely, by using the matrix $\bar {I^\alpha}_{\nu}=\mbox{diag} (1,1,0,0)$, they are exactly written as $\bar g_{\mu\nu} = g_{\mu\alpha} \bar {I^\alpha}_{\nu}$ and $\bar Z_{\mu\nu}=Z_{\mu\alpha} \bar {I^\alpha}_{\nu}$.

%

Varying the $\calL_2 $, $\calL_3 $ and $\calL_4 $ parts of the action with respect to $g^{\mu\nu}$ yields the followings
\begin{eqnarray}
&&\frac{1}{\sqrt{-g}}\frac{\delta}{\delta g^{\mu\nu}}\left(\int d^4x \sqrt{-g}\calL_2 \right)
= -\frac{1}{2}\calL_2 g_{\mu\nu}+\left(-3+[W]\right) \mathcal{W}_{\mu\nu} -\frac{1}{2}Z_{\mu\nu},
\label{VaryL2}\\
&&\frac{1}{\sqrt{-g}}\frac{\delta}{\delta g^{\mu\nu}}\left(\int d^4x \sqrt{-g}\calL_3 \right)
\nonumber\\
&&\qquad
= -\frac{1}{2}\calL_3 g_{\mu\nu}
+\left(-3+2[W] -\frac{1}{2}[W]^2+\frac{1}{2}[Z]\right) \mathcal{W}_{\mu\nu} 
+\left(-1+\frac{1}{2}[W]\right)Z_{\mu\nu}
-\frac{1}{3}\left({\cal W} Z\right)_{\mu\nu},\nonumber\\
&&
\label{VaryL3}\\
&&\frac{1}{\sqrt{-g}}\frac{\delta}{\delta g^{\mu\nu}}\left(\int d^4x \sqrt{-g}\calL_4 \right)
\nonumber\\
&&\qquad
= -\frac{1}{2}\calL_4 g_{\mu\nu}+\left( -1+[W]-\frac{1}{2}[W]^2+\frac{1}{6}[W]^3+\frac{1}{2}[Z]-\frac{1}{2}[W][Z]+\frac{1}{3}[WZ]\right) \mathcal{W}_{\mu\nu}
\nonumber\\
&&\qquad\qquad
+\left(- \frac{1}{2}+\frac{1}{2}[W]-\frac{1}{4}[W]^2+\frac{1}{4}[Z]\right)Z_{\mu\nu}
+\left(-\frac{1}{3}+\frac{1}{3}[W]\right)\left({\cal W} Z\right)_{\mu\nu}
-\frac{1}{4}Z_{\mu\alpha}Z^\alpha_{\ \nu}.
\label{VaryL4}
\end{eqnarray}
In order to investigate the difference between non-linear massive gravity and GR, 
we regard the mass terms in nonlinear massive gravity as effective energy-momentum sources in GR. From this point of view, (\ref{VaryL2}), (\ref{VaryL3}) and (\ref{VaryL4}) are, then, related to the effective energy-momentum tensor ${T^{\text{eff}}_{\mu\nu}}$ as
\begin{eqnarray}
T^{\text{eff}}_{\mu\nu}= - \frac{2}{\sqrt{-g}}\frac{\delta S_{mass}}{\delta g^{\mu\nu}}.
\end{eqnarray}
As calculated in Appendix~\ref{calmass}, each component of ${T^{\text{eff}}}_{\mu\nu}$  is written as
\begin{eqnarray}
&&T^{\text{eff}}_{mn}=-2M_{Pl}^2 m_g^2\biggl(\biggl[
\Bigl\{-3 +3X_{\pm}e^{-E}-\frac{1}{2}X_{\pm}^2e^{-2E}\nonumber\\
&&\qquad\qquad\qquad\qquad
+\alpha_3\left( -2+3X_\pm e^{-E}-X_\pm^2 e^{-2E}\right)
-\frac{\alpha_4}{2}\left(1-X_\pm e^{-E}\right)^2\Bigr\} \nonumber\\
&&\qquad
+{\mathbb{W}}_T\left\{\left(\frac{3}{2}-X_{\pm}e^{-E}\right) 
+\alpha_3\left(\frac{3}{2}-2X_\pm e^{-E}+\frac{1}{2}X_\pm^2 e^{-2E}\right)
+\frac{\alpha_4}{2} \left(1-X_\pm e^{-E}\right)^2\right\}\biggr] g_{mn} \nonumber\\
&&\qquad
+\left\{\left(-3+2X_\pm e^{-E}\right) 
+\alpha_3\left(-3+4X_\pm e^{-E}-X_\pm^2 e^{-2E}\right)
-\alpha_4\left(1-X_\pm e^{-E}\right)^2 \right\}{\cal W}_{mn}\biggr) ,
\label{Tmn}\\
&&T^{\text{eff}}_{ij}=-2M_{Pl}^2 m_g^2\biggl[
\left\{-3+\frac{3}{2}X_\pm e^{-E}+\alpha_3\left(-2+\frac{3}{2}X_\pm e^{-E}\right)
+\frac{\alpha_4}{2} \left(-1+X_\pm e^{-E}\right)\right\} \nonumber\\
&&\qquad\qquad\qquad\qquad
+{\mathbb{W}}_T\left\{\frac{3}{2}-\frac{1}{2}X_\pm e^{-E}+\alpha_3\left( \frac{3}{2}-X_\pm e^{-E} \right)+\frac{\alpha_4}{2} \left(1-X_\pm e^{-E}\right)\right\} \nonumber\\
&&\qquad\qquad\qquad\qquad
+\sqrt{\det (\mathbb{Z})}\left\{-\frac{1}{2} +\alpha_3\left(-1+\frac{1}{2}X_\pm e^{-E}\right)
+\frac{\alpha_4}{2} \left(-1+X_\pm e^{-E}\right)\right\} \biggr] g_{ij},
\label{Tij}
\end{eqnarray}
where $m,n = t, r$ and $i,j= \theta, \psi$ and the other components are zero.

Interestingly, for $E=0$, by using eq.(\ref{f}) $T^{\text{eff}}_{mn}$ is reduced to $-M_{Pl}^2\Lambda_{\text{eff}} g_{mn}$, namely there is no effective contribution except for the cosmological constant. 
Then, the form of $T^{\text{eff}}_{ij}$ can be fixed by the conservation law. 
The conservation law is obtained by take the divergence of eq.(\ref{gravitationaleq});
\begin{eqnarray}
\nabla^\mu T^{\text{eff}}_{\mu\nu} =0.
\end{eqnarray}
While for $\nu=i$ $(i=\theta,\psi)$ the conservation law becomes trivial because of the symmetry,
for $\nu= t$ and $\nu=r$ it gives the relation between $T^{\text{eff}}_{mn}$ and $T^{\text{eff}}_{ij}$. 
When $T^{\text{eff}}_{mn}=0$, in the both case for $\nu= t$ and $\nu=r$, it becomes the same constraint equation;
\begin{eqnarray}
0&\eq&\left( T_{ij}|_{E=0}+M_{Pl}^2\Lambda_{\text{eff}} g_{ij} \right) \delta^{ij}\nonumber\\
&\eq&-2M_{Pl}^2 m_g^2\left[-3+X_\pm +\alpha_3\left(-3+2X_\pm\right)+\alpha_4\left(-1+X_\pm\right)\right]
\left(X_\pm-{\mathbb{W}}_T+\frac{\sqrt{\det (\mathbb{Z})}}{X_\pm}\right) ,
\end{eqnarray}
where we use eq.(\ref{f}) and (\ref{Lambdaeff}). 
Since $T_{ij}$ is proportional to $g_{ij}$, this obviously results in 
\begin{eqnarray}
T_{ij}|_{E=0}=- M_{Pl}^2\Lambda_{\text{eff}} g_{ij}.
\end{eqnarray}
Therefore, for $E=0$, the contribution from the mass term is completely equal to the cosmological constant and its solution must be the same as the gauge fixed one of GR with the cosmological constant, i.e., de Sitter-Schwarzschild with the following gauge choice\footnote{
With the different gauge choice, the de Sitter-Schwarzschild solution was obtained in \cite{Koyama:2011xz,Koyama:2011yg,Comelli:2011wq,Nieuwenhuizen:2011sq}.
}:
\begin{eqnarray}
&&E=0, \\
&&\beta^2 =\frac{1-Kr^2}{a^2} \left(e^{2\Phi}-\frac{\dot a^2}{|K|}\right)\left(e^{-2\Psi}-1\right).
\end{eqnarray}
We linearly expand $T^{\text{eff}}_{\mu\nu}$ since in subsection~\ref{linearanalysis} we 
analyze the linearized equations.  
Assuming $\Phi, \beta, \Psi$ and $E$ are all the same order as a small parameter $\epsilon$ and linearly expanding (\ref{VaryL2}),  $T^{\text{eff}}_{\mu\nu}$ can be expanded as
\begin{eqnarray}
&&T^{\text{eff}}_{rr}+M_{Pl}^2\Lambda_{\text{eff}} g_{rr} 
\simeq
-2M_{Pl}^2 m_g^2 C_\pm
\left(1-\frac{\dot a}{\sqrt{|K|}}\right)E g_{rr}, \\
&&T^{\text{eff}}_{ij}+M_{Pl}^2\Lambda_{\text{eff}} g_{ij}\simeq 
- M_{Pl}^2m_g^2
C_\pm\left(1-\frac{\dot a}{\sqrt{|K|}}\right)\left(E+\Psi\right) g_{ij},\\
&&C_\pm \equiv \left\{\left(-3+X_{\pm}\right) 
+\alpha_3\left(-3+2X_\pm\right)
+\alpha_4\left(-1+X_\pm\right)\right\}X_\pm , \label{Cpm}
\end{eqnarray}
and the other components of $\left(T^{\text{eff}}_{\mu\nu}+M_{Pl}^2\Lambda_{\text{eff}} g_{\mu\nu}\right)$ are zero.

\subsection{Contribution from the Einstein-Hilbert term}
\label{EHcontribution}

In this subsection, we show the linear contribution from the Einstein-Hilbert term. 
Since the form of the Einstein-Hilbert term is the same as that in GR, 
the linearized parts of the contribution from the Einstein-Hilbert term must be written with the gauge invariant combinations of GR. 
Therefore, it is convenient to define new variables $A$ and $B$ as
\begin{eqnarray}
&&A \equiv \Psi - \left(\frac{a^2 r^2}{1-Kr^2}H^2\right) \Phi - \left(\frac{a^2}{1-Kr^2}\right)rH\beta + \left(\frac{a^2 r^2}{1-Kr^2}\right)H \dot{E} - \frac{E}{1-Kr^2}-rE', 
\label{defA}\\
&&B \equiv \frac{1}{H}\dot{\Psi}- \frac{\dot{H}}{H^2}\Psi + r\Phi' +\left(\frac{Kr^2}{1-Kr^2}\right)\Phi - \frac{r}{H}\dot{E}' - \frac{\dot{E}}{H(1-Kr^2)}+\frac{\dot{H}}{H^2}rE' +\frac{\dot{H}}{H^2}\frac{E}{1-Kr^2},\nonumber\\
&&\label{defB}
\end{eqnarray}
where prime and dot denote derivative with respect to radial and time coordinate respectively, and which are gauge invariant variables in GR. (See Appendix~\ref{App2}.) 
We stress that in massive gravity, since there are no gauge degrees of freedom, these variables $A$ and $B$ are not gauge invariant variables but  are just the linear combinations of the physical variables $\Phi$, $\beta$, $\Psi$ and $E$. 
These variables are useful to write the linear contributions from the Einstein-Hilbert term and to discuss the difference between massive gravity and GR.

Equipped with these two variables, we have
\begin{eqnarray}
R_{00}-\frac{1}{2}R g_{00}+ \Lambda_{\text{eff}} g_{00} &\eq& \frac{2(1-Kr^2)}{a^2r^2}A + \frac{2 (1-Kr^2)}{a^2r}A' + 2H^2 B + \calO(\epsilon^2),\nonumber  \\
R_{0r}-\frac{1}{2}R g_{0r}+ \Lambda_{\text{eff}} g_{0r} &\eq&\frac{2}{r}\left(\frac{\dot{H}}{H}A+HB \right)+\calO(\epsilon^2) ,\nonumber\\
R_{rr}-\frac{1}{2}R g_{rr}+ \Lambda_{\text{eff}} g_{rr} &\eq& \frac{2}{r^2}\left[\left(\frac{\dot{H}}{H^2}-1 \right)A - \frac{1}{H}\left(\dot{A} - HB \right) \right]+ \calO(\epsilon^2) ,
\label{Gmn}\\
R_{ij}-\frac{1}{2}R g_{ij}+ \Lambda_{\text{eff}} g_{ij} &\eq&
\biggl[r(1-Kr^2)B'- a^2 r^2(\dot{H}B + H\dot{B})-(2Kr^2+3a^2r^2H^2)B \nonumber\\
&&- \frac{r(1-Kr^2)}{H}\dot{A}'+ \frac{Kr^2}{H}\dot{A}-r(1-Kr^2)\left(1- \frac{\dot{H}}{H^2}\right)A'   \nonumber\\
&& + Kr^2\left( 1- \frac{\dot{H}}{H^2}\right)A+\calO(\epsilon^2)\biggr]\frac{1}{a^2r^2}g_{ij}.\nonumber
\end{eqnarray}

\subsection{Linear analysis}
\label{linearanalysis}

In order to investigate the properties of massive gravity theory, 
we focus on the week gravity case and solve the linearized equations. 
Instead of the four variables $\Phi$, $\beta$, $\Psi$ and $E$, we use the variables $A$, $B$, $\Psi$ and $E$ as independent variables.
It is useful because equations become simpler and because $A$ and $B$ are gauge invariant combinations in GR.
The contributions of the linear order is obtained in \S~\ref{masscontribution} and \S~\ref{EHcontribution} and the linearized equations of motion become
\begin{align}
&\frac{2(1-Kr^2)}{a^2r^2}A + \frac{2 (1-Kr^2)}{a^2r}A' + 2H^2 B =0 ,\label{EOM1}\\
&\frac{2}{r}\left(\frac{\dot{H}}{H}A+HB \right)=0, \label{EOM2}\\
&\frac{2}{r^2}\left[\left(\frac{\dot{H}}{H^2}-1 \right)A - \frac{1}{H}\left(\dot{A} - HB \right) \right] = -2 m_g^2 C_\pm \left(\frac{a^2}{1-Kr^2} \right)\left( 1- \frac{aH}{\sqrt{|K|}}\right)E, \label{EOM3}\\
&r(1-Kr^2)B'- a^2 r^2(\dot{H}B + H\dot{B})-(2Kr^2+3a^2r^2H^2)B - \frac{r(1-Kr^2)}{H}\dot{A}'\nonumber\\
&\quad \quad \quad  + \frac{Kr^2}{H}\dot{A}-r(1-Kr^2)\left(1- \frac{\dot{H}}{H^2}\right)A' = -m_g^2 C_\pm a^2r^2\left(1- \frac{aH}{\sqrt{|K|}} \right)(\Psi + E) . \label{EOM4}
\end{align}
Recall that $\dot{H} = K/a^2$, hence (\ref{EOM2}) gives
\begin{eqnarray}
B = - \frac{K}{a^2H^2}A.
\label{BandA}
\end{eqnarray}
Substituting this into (\ref{EOM1}) one obtains
\begin{eqnarray}
A' = -\frac{(1-3Kr^2)}{r(1-Kr^2)}A.  \label{EOMA}
\end{eqnarray}
With these two relations, the last two equations of motion (\ref{EOM3}) and (\ref{EOM4}) become
\begin{eqnarray}
-\frac{2}{H^2r^2}\left(H^2A + H\dot{A}\right)&=&-2m_g^2C_\pm\left( \frac{a^2}{1-Kr^2}\right)\left( 1- \frac{aH}{\sqrt{|K|}}\right)E ,
\label{equE}\\
\frac{(1-Kr^2)}{H^2}\left(H^2 A + H\dot{A} \right) &=& -m_g^2C_\pm a^2r^2\left(1-\frac{aH}{\sqrt{|K|}} \right)\left(\Psi + E \right),
\end{eqnarray}
which yield
\begin{eqnarray}
\Psi=-2E. \nonumber \label{psiandE}
\end{eqnarray}
Solving (\ref{EOMA}) and substituting the solution into eqs.(\ref{BandA}), (\ref{equE}) and 
(\ref{psiandE}), we have 
\begin{eqnarray}
A&=&\frac{1}{8 \pi M_{Pl}^2}\frac{M(t)}{ar(1-Kr^2)}, \\
B&=&- \frac{1}{8 \pi M_{Pl}^2}\frac{K}{a^2H^2}\frac{M(t)}{ar(1-Kr^2)}, \\
E&=& \frac{1}{8 \pi M_{Pl}^2}\frac{1}{m_g^2 C_\pm}\left(1-\frac{aH}{\sqrt{|K|}} \right)^{-1} \frac{\dot{M}(t)}{a^3r^3}, \\
\Psi &=&- \frac{1}{8 \pi M_{Pl}^2}\frac{2}{m_g^2 C_\pm}\left(1-\frac{aH}{\sqrt{|K|}} \right)^{-1} \frac{\dot{M}(t)}{a^3r^3},
\end{eqnarray}
where $M(t)$ is an arbitrary time-dependent function.
The solution has one time-dependent functional parameter, which is very different from that in GR.
Only when $M(t)$ is a constant, the solution is the same as the linearized de Sitter-Schwarzschild solution and 
$M(t)$ is corresponding to the mass of a star (or a black hole). 
Then, $E$ and $\Psi$ are zero, which is consistent with the non-linear result in \S~(\ref{masscontribution}).

\section{ Summary and Discussion}
\label{Sum}

In this paper, we have analyzed the spherically symmetric solutions with asymptotically open FLRW solution in non-linear massive gravity. 
In \S~\ref{masscontribution}, we have presented the non-linear form of the effective energy-momentum tensor which comes from the mass term. 
Only in the special case where $E=0$, the effective energy-momentum tensor is completely the same as that of cosmological constant. 
Otherwise, other effects show up. 
This gives a possibility to discriminate the massive gravity theory from GR. 

In \S~\ref{linearanalysis}, we have solved the gravitational equations of motion in the vacuum in linear order of metric perturbation and obtained the result that the solution depends on a time-dependent parameter. 
In GR, the parameter is a constant and corresponding to the mass of a star. 
 In general, massive gravity should have five propagating degrees of freedom, but kinetic terms of the helicity-0 and helicity-1 modes mysteriously vanish in some background. Nevertheless, these mode could still leave their trace in the form of static potential. This is probably the reason that while in the analysis of gravitational propagation~\cite{Gumrukcuoglu:2011zh} the helicity-0 and helicity-1 modes do not show up  and the difference from GR in linear order appears only in the mass of helicity-2 modes, 
in our analysis of  the static potential we can see the large difference from GR even in linear order.

Our result means that in massive gravity Birkhoff's theorem does not hold true 
and suggests that energy can be gravitationally emitted even in spherically-symmetric configuration. 
The existence of the helicity-0 mode is expected from this result because the spherically-symmetric gravitational wave is probably related to the helicity-0 mode. 
This gives a hint for the question if the propagation of the helicity-0 mode accidentally disappears only on open FLRW solution. 
The fact that the parameter only depends on time but not on the radial coordinate implies superluminous behavior.
In a construction of a concrete solution,
we choose the parameter at some point on each time-constant hypersurface, such as the surface of a star, and then, metric at other points on the same time-constant hypersurface is fixed.  
This means that
its information propagates to all points on the same time-constant hypersurface, namely 
the speed of the propagation is infinite. 
Infinite speed of propagation can be realized when the coefficient of the kinetic term (time-derivative term) in the action vanishes. 
Therefore, we can conclude that our result shows the sign of superluminous (infinite-speed) behavior of the helicity-0 mode in the self-accelerating background.

If the matter is coupled with the superluminous helicity-0 mode, the infinite-speed propagation makes the situation hazardous, 
although we do not show the coupling of gravity with matter in this paper. 
The superluminous helicity-0 mode conveys the information of matter to infinity for an instant. 
It can be observed as the violation of the energy conservation. 
In order to avoid the pathology in massive gravity, 
decoupling between matters and the helicity-0 mode (and probably also helicity-1 modes) is needed. 

In order to see the disappearance of the kinetic term directly, 
it is better to analyze the perturbation on the de Sitter-Schwarzschild solution and to take the limit that the mass parameter of the solution goes to zero. 
If our expectation that the kinetic term of the helicity-0 mode accidentally disappears only on open FLRW solution is correct, the helicity-0 mode can propagate on the de Sitter-Schwarzschild background. 
Then if we take the massless limit of black hole mass, we can check the disappearance of the kinetic term. 
Perturbation on the de Sitter-Schwarzschild background will be discussed in an upcoming work.

\section*{Acknowledgement}
We would like to thank Shinji  Mukohyama and A. Emir G\"umr\"uk\c{c}\"uo\u{g}lu for useful discussions and comments. C.-I C. and K.~I. acknowledge supports by Taiwan National Science Council under Project No.~NSC100-2119-M002-025. P.~C.~ is supported by Taiwan National Science Council under
Project No.~NSC97-2112-M-002-026-MY3, by TaiwanÕs National Center
for Theoretical Sciences (NCTS), and by US Department of Energy under
Contract No.~DE-AC03-76SF00515.

\appendix

\section{Calculation of contribution from mass term}
\label{calmass}

In this appendix, we show the brief calculations about the mass term.
On the metric~(\ref{metric}), the concrete forms of $[W]$, $[Z]$, $[WZ]$ and $[Z^2]$ become 
\begin{eqnarray}
&&[W]={\mathbb{W}}_T +X_{\pm} 2e^{-E},\\
&&[Z]=  {\mathbb{W}}_T^2- 2\sqrt{\det (\mathbb{Z})} +X_\pm^2 2e^{-2E},\\
&&[WZ]={\mathbb{W}}_T^3 -3{\mathbb{W}}_T \sqrt{\det (\mathbb{Z})}+2X_\pm^3 e^{-3E}, \\
&&[Z^2]={\mathbb{W}}_T^4-4{\mathbb{W}}_T^2\sqrt{\det (\mathbb{Z})} +2 \det (\mathbb{Z}) +2 X_\pm^4 e^{-4E}.
\end{eqnarray}
Substituting them into eqs.(\ref{L_2})-(\ref{L_4}), we have
\begin{eqnarray}
&&\frac{1}{2}{\cal L}_2= 3 -3X_{\pm}e^{-E}+\frac{1}{2}X_{\pm}^2e^{-2E}
-{\mathbb{W}}_T\left(\frac{3}{2}-X_{\pm}e^{-E}\right)
+\frac{1}{2}\sqrt{\det (\mathbb{Z})},\\
&&\frac{1}{2}{\cal L}_3=2-3X_\pm e^{-E}+X_\pm^2 e^{-2E}+
\left(-\frac{3}{2}+2X_\pm e^{-E}-\frac{1}{2}X_\pm^2 e^{-2E}\right){\mathbb{W}}_T
+\left(1-X_\pm e^{-E} \right)\sqrt{\det (\mathbb{Z})} ,\nonumber\\
&&\\
&&\frac{1}{2}{\cal L}_4= \frac{1}{2} \left(1-X_\pm e^{-E}\right)^2 
\left(1-{\mathbb{W}}_T + \sqrt{\det (\mathbb{Z})}\right) . 
\end{eqnarray}
Then, the variation of ${\cal L}_2$, ${\cal L}_3$ and ${\cal L}_4$ of the action with respect to $g^{\mu\nu}$ can be written as 
\begin{eqnarray}
&&\frac{1}{\sqrt{-g}}\frac{\delta}{\delta g^{\mu\nu}}\left(\int d^4x \sqrt{-g}\calL_2 \right)
\nonumber\\
&&\qquad 
=-\frac{1}{2}\calL_2 g_{\mu\nu}+\left(-3+2X_\pm e^{-E}+{\mathbb{W}}_T\right) \mathcal{W}_{\mu\nu} -\frac{1}{2}Z_{\mu\nu}\nonumber\\
&&\qquad
=\left[-3 +3X_{\pm}e^{-E}-\frac{1}{2}X_{\pm}^2e^{-2E}
+{\mathbb{W}}_T\left(\frac{3}{2}-X_{\pm}e^{-E}\right) \right]{\bar g}_{\mu\nu}
+\left(-3+2X_\pm e^{-E}\right){\bar {\cal W}}_{\mu\nu}\nonumber\\
&&\qquad\qquad
\left(-3+\frac{3}{2}X_\pm e^{-E} 
+\left(\frac{3}{2}-\frac{1}{2}X_\pm e^{-E}\right){\mathbb{W}}_T -\frac{1}{2}\sqrt{\det (\mathbb{Z})} \right)
\left(g_{\mu\nu}-{\bar g}_{\mu\nu}\right) ,
\label{AppL2}
\end{eqnarray}
\begin{eqnarray}
&&\frac{1}{\sqrt{-g}}\frac{\delta}{\delta g^{\mu\nu}}\left(\int d^4x \sqrt{-g}\calL_3 \right)
\nonumber\\
&&\qquad
= -\frac{1}{2}\calL_3 g_{\mu\nu}
+\left(-3+4X_\pm e^{-E}-X_\pm^2 e^{-2E}+(2-2X_\pm e^{-E}){\mathbb{W}}_T-\sqrt{\det (\mathbb{Z})}\right) \mathcal{W}_{\mu\nu} 
\nonumber\\
&&\qquad\qquad
+\left(-1+X_\pm e^{-E}+\frac{1}{2}{\mathbb{W}}_T\right)Z_{\mu\nu}-\frac{1}{3}\left({\cal W} Z\right)_{\mu\nu}\nonumber\\
&&\qquad
=\left[ -2+3X_\pm e^{-E}-X_\pm^2 e^{-2E}+
\left(\frac{3}{2}-2X_\pm e^{-E}+\frac{1}{2}X_\pm^2 e^{-2E}\right){\mathbb{W}}_T \right]{\bar g}_{\mu\nu} \nonumber\\
&&\qquad\qquad
+\left(-3+4X_\pm e^{-E}-X_\pm^2 e^{-2E}\right){\bar {\cal W}}_{\mu\nu}
+\biggl[-2+\frac{3}{2}X_\pm e^{-E}
+\left( \frac{3}{2}-X_\pm e^{-E} \right){\mathbb{W}}_T\nonumber\\
&&\qquad\qquad\qquad\qquad\qquad\qquad\qquad
+\left(-1+\frac{1}{2}X_\pm e^{-E}\right) \sqrt{\det (\mathbb{Z})} \biggr]
\left(g_{\mu\nu}-{\bar g}_{\mu\nu}\right), \label{AppL3}
\end{eqnarray}
\begin{eqnarray}
&&\frac{1}{\sqrt{-g}}\frac{\delta}{\delta g^{\mu\nu}}\left(\int d^4x \sqrt{-g}\calL_4 \right)
\nonumber\\
&&\qquad
= -\frac{1}{2}\calL_4 g_{\mu\nu}+
\Bigl\{-1+2X_\pm e^{-E}-X_\pm^2 e^{-2E}
+{\mathbb{W}}_T\left(1-2X_\pm e^{-E}+X_\pm^2 e^{-2E}\right) 
\nonumber\\
&&\qquad\qquad\qquad\qquad\qquad\qquad\qquad\qquad\qquad\qquad\qquad\qquad
+\left(-1+2X_\pm e^{-E}\right) \sqrt{\det (\mathbb{Z})}
\Bigr\}\mathcal{W}_{\mu\nu}
\nonumber\\
&&\qquad\qquad
+\left\{- \frac{1}{2}+X_\pm e^{-E}-\frac{1}{2}X_\pm^2 e^{-2E}
+{\mathbb{W}}_T\left(\frac{1}{2}-X_\pm e^{-E}\right)
-\frac{1}{2}\sqrt{\det (\mathbb{Z})}
\right\}Z_{\mu\nu}
\nonumber\\
&&\qquad\qquad
+\left(-\frac{1}{3}+\frac{2}{3}X_\pm e^{-E}+\frac{1}{3}{\mathbb{W}}_T\right)\left({\cal W} Z\right)_{\mu\nu}
-\frac{1}{4}Z_{\mu\alpha}Z^\alpha_{\ \nu}\nonumber\\
&&\qquad
=\left(1-X_\pm e^{-E}\right)^2 \biggl(-\frac{1}{2}  
\left(1-{\mathbb{W}}_T \right){\bar g}_{\mu\nu} -{\bar {\cal W}} _{\mu\nu}\biggr)\nonumber\\
&&\qquad\qquad
+\left(1-X_\pm e^{-E}\right)
\left(-1+{\mathbb{W}}_T -\sqrt{\det (\mathbb{Z})}\right) \left(g_{\mu\nu}-{\bar g}_{\mu\nu}\right),
\label{AppL4}
\end{eqnarray}
where ${\bar g}_{\mu\nu}$ is defined under eq.(\ref{Z2}) and ${\bar {\cal W}} _{\mu\nu}={\cal W} _{\mu\alpha} {{\bar I}}^{\alpha}_{\nu}$.
The effective energy-momentum tensor $T^{\text{eff}}_{\mu\nu}$ can be 
written with the linear combination of eqs.(\ref{AppL2})-(\ref{AppL4}) as eqs.(\ref{Tmn}) and (\ref{Tij}) and
linearly expanded as 
\begin{eqnarray}
&&T^{\text{eff}}_{mn}+M_{Pl}^2\Lambda_{\text{eff}}g_{mn} 
\simeq
-2M_{Pl}^2 m_g^2\frac{C_\pm}{X_\pm} E
\left(X_\pm g_{mn}-{\mathbb{W}}_T^{(0)} g_{mn} +2{\cal W}_{mn}^{(0)}\right), \\
&&T^{\text{eff}}_{ij}+M_{Pl}^2\Lambda_{\text{eff}} g_{ij}\simeq 
-M_{Pl}^2m_g^2
C_\pm\left(1-\frac{\dot a}{\sqrt{|K|}}\right)\left(E+\Psi\right) g_{ij},
\end{eqnarray}
where $C_\pm$ is defined in (\ref{Cpm}) 
Background values of ${\mathbb{W}}_T$
and ${\cal W}_{mn}$ are, respectively, 
\begin{eqnarray}
&&{\mathbb{W}}_T^{(0)}=X_\pm \left(\frac{\dot a}{\sqrt{|K|}}+1\right),\\
&&{\cal W}_{mn}^{(0)}= \left(
  \begin{array}{cc}
 \frac{  X_\pm }{2} \frac{\dot a}{\sqrt{|K|}} g_{tt}^{(0)} & 0   \\
      0& \frac{  X_\pm }{2}  g_{rr}^{(0)}
  \end{array}
\right).
\end{eqnarray}
From these, we know the only rr-component of linearized $\left(T_{mn}-\Lambda g_{mn}\right)$ becomes non-zero:
\begin{eqnarray}
T^{\text{eff}}_{rr}+M_{Pl}^2\Lambda_{\text{eff}}g_{rr} 
\simeq
-2M_{Pl}^2 m_g^2 
C_\pm \left(1-\frac{\dot a}{\sqrt{|K|}}\right)E g_{rr}.
\end{eqnarray}

\section{Derivation of contribution from the Einstein-Hilbert term}
\label{App2}

In this subsection, we derive the linear contribution from the Einstein-Hilbert term.
Because the contribution from the Einstein-Hilbert term should be the same as that of GR, 
we can use the same technique as in GR. 
In concrete terms, we construct the gauge invariant variables and 
write the perturbative terms with them.
Since the background we consider is accelerated by the effective energy momentum tensor, 
we add the same contribution in GR by hand, namely we consider the perturbation of
\begin{eqnarray}
R_{\mu\nu}-\frac{1}{2}R g_{\mu\nu}+\Lambda_{\text{eff}} g_{\mu\nu}.
\label{GRGmn}
\end{eqnarray}
In order to keep spherical symmetry, we would only consider diffeomorphisms in the temporal and radial coordinate: 
\begin{eqnarray}
x^\mu \rightarrow x'^\mu = x^\mu + \xi^\mu, \quad\quad\quad\quad \xi^\mu = (\xi^t, \xi^r, 0, 0)\equiv(\zeta, \xi, 0, 0).
\end{eqnarray}
Under such gauge transformation, the metric perturbations transform as
\begin{eqnarray} 
\begin{cases}
\displaystyle \Phi  \rightarrow  \Phi- \dot{\zeta}, \\
\displaystyle \beta \rightarrow  \beta + \left(\frac{1-Kr^2}{a^2} \right)\zeta' - \dot{\xi},\\
\displaystyle \Psi  \rightarrow  \Psi - H\zeta - \left(\frac{Kr}{1-Kr^2} \right)\xi-\xi' ,\\
\displaystyle E     \rightarrow  E - H\zeta - \frac{1}{r}\xi,
\end{cases}
\end{eqnarray}
where prime and dot denote derivative with respect to the radial and time coordinate respectively. 
Then, the gauge invariant variables $A$ and $B$ are written as eqs.(\ref{defA}) and (\ref{defB}). 
Equipped with these two variables, the perturbation of (\ref{GRGmn}) can be expressed as eq.(\ref{Gmn}).

\bibliographystyle{JHEP}
\bibliography{MassiveGravity}
 
\end{document}